\documentclass[prl,aps,showpacs,twocolumn]{revtex4}
\usepackage{graphicx}
\begin{document}

\title{Random matrices, symmetries, and many-body states}

\author{Calvin W. Johnson}
\affiliation{Department of Physics, San Diego State University,
5500 Campanile Drive, San Diego, CA 92182-1233}


\begin{abstract}
All nuclei with even numbers of protons and of neutrons have ground states with 
zero angular momentum. This is  ascribed to the 
pairing force between nucleons, but  simulations with 
random  interactions suggest 
a much broader many-body phenomenon. In this Letter I 
project out random Hermitian matrices that have good quantum numbers and, 
computing the width of the Hamiltonian in subspaces, 
find ground states dominated by 
low quantum numbers, e.g. $J = 0$.  Furthermore I find  odd-$Z$, odd-$N$ systems 
with isospin conservation have relatively fewer $J=0$ ground states. 
\end{abstract}

\maketitle

Most quantum many-body systems cannot be exactly solved, even numerically.  
Because generic many-body systems 
are complex and classically chaotic, one way to model the spectrum is through random 
matrices \cite{RM}. 
These matrices must be Hermitian, of course, but there are no other
\textit{exact} symmetries.

Real many-body system often have non-trivial symmetries such as rotational invariance and isospin invariance,
which give rise to states with exact quantum numbers such as angular momentum $J$, $J_z$, and isospin $T$, $T_z$.  But because 
the above random matrices do not have such symmetries, investigators tended to consider 
statistical properties of states with the same quantum numbers. 

In contrast, the ordering of different quantum numbers in spectra was 
associated with details of the interaction. For 
example, the fact that nuclei with an even number $Z$ of protons and an even number $N$ 
of neutrons always have ground states with angular momentum $J=0$, while odd-$Z$, odd-$N$ 
nuclei frequently do not, was attributed to the pairing interaction \cite{BM,simple}.

It was therefore  a shock to discover that rotationally invariant but 
otherwise random two-body Hamiltonians  
tend to yield ground states with $J=0$, 
just like `realistic' interactions, even though such states are a small fraction 
of the total space\cite{JBD98,ZV04,ZAY04}. This phenomenon is robust, insenstive to details 
of the distribution of matrix elements \cite{Jo99}, occurs not only for 
fermions but also for bosons \cite{BF00}, 
and is relatively insensitive to the particle rank of the interaction \cite{Vo08}.
 Over the past decade there have been many papers proposing explanations. 
As the distribution of many-body systems with two-body interactions tend to have a 
Gaussian distribution of states \cite{MF75}, a number of authors have focused on 
widths \cite{BFP99,PW04}, while others have statistically averaged in a single $j$-shell 
the coupling of
multiple angular momenta \cite{MVZ00}. As a 
recent Letter stated, `the simple question of symmetry and chaos asks for a simple 
answer which is still missing \cite{Vo08}.'

In this paper I return to random matrices and impose  symmetries,  first U(1) then 
SU(2).  I show explicitly how combining `internal' degrees of freedom with 
projection of good quantum numbers leads naturally to subspaces with 
small quantum numbers having the greatest widths, and thus dominating the ground state. 
This simple picture applies with equal ease to both fermionic and bosonic systems, 
and aside from subspace dimensions is independent of the detailed microphysics, 
helping to explain the robustness of the phenomenon. 
And considering two simultaneous SU(2) symmetries, angular momentum and isospin,  
I find conservation of isospin in odd-$Z$, odd-$N$ system 
changes dimensions of subspaces in such a way as to decrease the fraction of $J=0$ ground states, a prediction confirmed with 
detailed simulations. 

I start with U(1) symmetry and
consider a wavefunction $\psi(\phi)$ which is periodic $\psi(\phi+2\pi) = \psi(\phi)$ and
which is an eigenstate of a general  eigenvalue equation:
\begin{equation}
\int_0^{2\pi} {H}(\phi,\phi^\prime)\psi(\phi^\prime) d\phi^\prime = E \psi(\phi)
\end{equation}
%(Note this integral equations includes derivatives and other common expressions). 
Without significant loss of generality I assume the wavefunction $\psi$ and the Hamiltonian ${H}(\phi,\phi^\prime)$ to 
be real. Hermiticity requires that $H(\phi,\phi^\prime) = H(\phi^\prime,\phi)$, while 
U(1) invariance suggests that ${H}$ can only depend on the 
\textit{difference} of angles: $ {H}(\phi,\phi^\prime)= F(\phi-\phi\prime)$.
Combining Hermiticity with periodicity leads to  $F(x) = F(2\pi-x)$.

Inasmuch as $F$ is a periodic function, I make a Fourier decomposition, 
keeping in mind that $F$ is real, % and the symmetry about $x = \pi$, 
\begin{equation}
F(x) = \frac{1}{2\pi} \sum_m h_m \cos ( mx)
\end{equation}
which can be inverted
\begin{equation}
h_m = \int_0^{\pi} 2\cos(mx) F(x) dx \label{fourier}
\end{equation}
so that the integral is only over unique values of $F(x)$. 

If $F(x)$ is a randomly distributed variable about $x=0$, 
with a variance $\bar{\sigma}^2$ independent of $x$, then 
on average the value of $h_m$ is zero and the variance is easily 
computed:
\begin{eqnarray}
\sigma^2(h_m) = \int_0^{\pi} 4\cos^2(mx) F^2(x) dx \nonumber \\
= \int_0^{\pi} 4\cos^2(mx) \bar{\sigma}^2  dx = 2 \pi (1+ \delta_{m,0}) \bar{\sigma}^2.
\label{U1variance}
\end{eqnarray}

If the only degree of freedom is  $\phi$, then
the $h_m$ are the eigenvalues of $H$, and $m$ both labels the 
solutions and their symmetry. But now suppose there are (discrete) 
internal degrees of freedom, for instance if one has a many-body system that has 
an overall U(1) symmetry. In that case $\mathbf{F}(x)$ is a \textit{matrix-valued} 
function of $x$, and $\mathbf{h}_m$ is a random symmetric matrix, with the dimensions 
of $\mathbf{F}$ counting internal degrees of freedom.

If the matrix elements of $\mathbf{F}(x)$ each have a variance of
 $\bar{\sigma}$ independent of $x$, then the variance of the individual matrix element of 
 $\mathbf{h}_m$ are still given by (\ref{U1variance}).
Hence the matrix for $m =0$ has twice the variance of $m >0$ and the ground state will 
be likely have $m = 0$. 

I can illuminate this by discretizing $\phi$. Then the symmetry forces a matrix of the 
form
\begin{equation}
\mathbf{H} = 
\left (
\begin{array}{ccccccc}
A & B & C & D & \ldots & C  & B \\
B & A & B & C & \ldots & D  & C \\
C & B & A & B & \ldots & E  & D  \\
\vdots & & & & & & \\
B & C & D & E & \ldots & B  & A 
\end{array}
\right ) \label{hdef}
\end{equation}
Adding `internal' degrees of freedom means that $A,B,C,\dots$ are now random symmetric 
matrices of the same dimension.  Numerical calculations with (\ref{hdef}) verify 
the accuracy of (\ref{fourier}) and (\ref{U1variance}) and the dominance of $m = 0$ quantum numbers 
for the ground state. 

Now  consider SU(2), rotational invariance.  
Using the angles
from spherical coordinates, the Hamiltonian is of 
the form ${H}(\theta^\prime \phi^\prime, 
\theta \phi)$, but imposing rotational invariance means $\hat{H}$ can only 
depend on the angle $\omega$ \textit{between} $\theta^\prime, \phi^\prime$ and
$\theta, \phi$ as given by 
$\cos \omega = \cos \theta \cos \theta^\prime + \sin \theta \sin \theta^\prime 
\cos( \phi - \phi^\prime)$. Then much like U(1)
\begin{equation}
{H}(\theta^\prime \phi^\prime, \theta \phi) = F( \omega),
\end{equation}
where $F(\omega) = F(\omega+2\pi)$ is a periodic function and, using Hermiticity 
(and assuming $H$ is real) $F(\omega) = F(2\pi-\omega)$ 
is symmetric with respect to $\omega = \pi$. 
Expanding
\begin{eqnarray}
F(\omega) = \sum_J h_J\frac{2J+1}{4\pi} P_J(\cos \omega) \nonumber \\
= \sum_J h_J \sum_{M = -1}^J Y_{JM}(\theta^\prime,\phi^\prime) Y^*_{JM}(\theta,\phi)
\end{eqnarray}
so clearly the $h_J$ are again the eigenvalues, with $Y_{JM}$ as eigenfunctions 
and with the eigenvalues independent of $N$, as one expects.  
%Note: I use $J$ to denote generic total angular momentum, which may be composed from both orbital and spin angular momenta. 

Once more I assume $\mathbf{F}(\omega)$ to be a matrix-valued function, and 
thus $\mathbf{h}_J$ to be a symmetric matrix given by
\begin{equation}
\mathbf{h}_J = 2\pi \int_0^\pi P_J(\cos \omega) \mathbf{F}(\omega) \sin \omega d\omega.
\end{equation}
As before, let $\bar{\sigma}$ be the variance of the matrix elements $\mathbf{F}$ independent
of $\omega$. 
Then I estimate the variance of the matrix elements of $\mathbf{h}_J$ 
\begin{equation}
\sigma^2_J = 4\pi^2 \bar{\sigma}^2 \int_0^\pi P_J^2(\cos \omega) \sin^2 \omega d\omega 
\label{SU2variance}
\end{equation}
%where the weighting factor $\sin^2 \omega$ arises because when I discretize, I assume the fluctuations in $\mathbf{F}$ should be binned  with respect to $\omega$ and not $\cos \omega$.  
Eqn.~(\ref{SU2variance}) can be computed numerically, and leaving off the 
factor $4\pi^2 \bar{\sigma}^2$, the  values are 
1.571, 0.393,  0.245, 0.178, 0.139  for $J = 0,1,2,3,4$ respectively. 

This suggests that in many-body system, subspaces with low-valued quantum numbers
will have larger widths. 
But in realistic, finite many-body calculations, subspaces with 
different $J$s have different dimensions. 
Furthermore, in the above argument each
$\mathbf{h}_J$ has independent random matrix elements, 
which typically has a semi-circular density of states \cite{RM}, 
yet for many-body systems with only two-body interactions the 
density of states tends towards a Gaussian\cite{MF75}. 

\begin{figure}  
\includegraphics [width = 7.5cm]{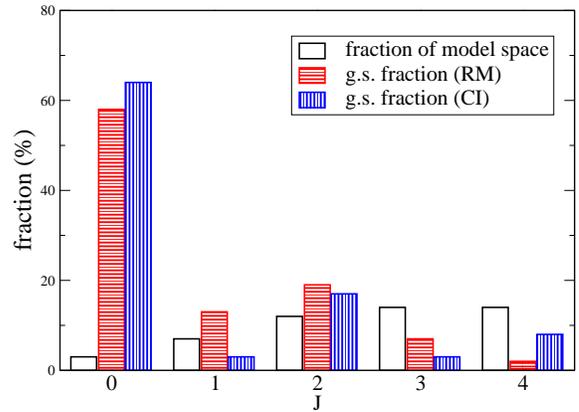}
\caption{\label{ca} (color online) 
Distribution of ground state quantum numbers
for eight neutrons in the $1p$-$0f$ shell. 
Empty bars are the fraction of states in the 
model space with a given $J$, horizontally striped (red) bars 
are the fraction of ground states with a given $J$ predicted 
by a random matrix (RM) model, and vertically striped (blue) bars 
are fraction of ground states with a given $J$ from 
configuration-interaction (CI) diagonalizations of an ensemble
of random two-body 
Hamiltonian in a shell-model basis. }
\end{figure}

I can approximately correct both deficiencies. First, following standard 
results on matrices with Gaussian-distributed matrix elements \cite{RM}, I let
\begin{equation}
\sigma_J^{\mathrm{eff}} = \sqrt{N_J} \sigma_J \label{scaledwidth}
\end{equation}
be the width of the subspace of states with angular momentum $J$. 
Then, for each $J$, I simply create $N_J$  energies via a random Gaussian distribution 
of width 
$\sigma_J^{\mathrm{eff}}$, and ultimately determine the fraction 
of ground states with angular momentum $J$. 

Finally, I compare against a variety of concrete simulations via configuration-interaction
calculations, that is, diagonalizing 
the Hamiltonian for fixed numbers of 
particles in finite single-particle spaces. 
Figure \ref{ca} shows the case of eight fermions (neutrons) in the 
$1p_{1/2}$-$1p_{3/2}$-$0f_{5/2}$-$0f_{7/2}$ shell-model space (which in nuclear physics 
corresponds to $^{48}$Ca with an inert $^{40}$Ca core), while 
Table \ref{singlespecies} considers two more cases, eight identical fermions in a $j=21/2$ shell, and six bosons in the
interacting boson model (IBM) \cite{IBM}. 
%The fermion cases were diagonalized using 
%the BIGSTICK code \cite{BIGSTICK} while the boson cases were computed using PHINT \cite{PHINT}.

For each of these many-body systems I take an 
ensemble of rotationally invariant, two-body but otherwise random interactions (typically
a few thousand cases), and tabulate the fraction $f_\mathrm{CI}$ of states that have a given 
angular momentum $J$.  This should be compared with the native fraction of states with 
that $J$ in each many-body space, $f_\mathrm{space} = N_J/N_\mathrm{tot}$ 
($N_\mathrm{tot}$ is the total dimension of the many-body space), and as noted originally the 
fraction with $J=0$ is dramatically enhanced.

\begin{table}
\caption{ Ground state quantum numbers for  single-species 
systems, comparing the percentage of ground states of a given $J$ 
for configuration-interaction simulations ($f_\mathrm{CI}$)
against the percentage computed using the simple random-matrix model 
($f_\mathrm{RM}$) described in the text. As input, $f_\mathrm{space}$ is the 
fraction of the states with a given $J$ in the CI model space. 
The two cases are 8 fermions in a $j=21/2$ shell, and the 
$N=7$ interacting 
boson model (IBM). 
\label{singlespecies}}
\begin{tabular}{|r|r|r|r|c|r|r|r|r|}
\cline{1-4} \cline{6-9}
\multicolumn{4}{|c|}{ $(21/2)^8$ } & & \multicolumn{4}{|c|}{ IBM, $N=7$ } \\
\cline{1-4} \cline{6-9}
$J$ &  $f_\mathrm{space}$ & $f_\mathrm{RM}$ &  $f_\mathrm{CI}$ &  &
$J$ &  $f_\mathrm{space}$ & $f_\mathrm{RM}$ &  $f_\mathrm{CI}$ \\
\cline{1-4} \cline{6-9}
0  & 0.4  & 33 & 55 &  & 0  & 11  & 81 & 55 \\
\cline{1-4} \cline{6-9}
1 &  0.5 & 0.2  & 0 & & 1 &  \multicolumn{3}{|c|}{ (no states)} \\
\cline{1-4} \cline{6-9}
2 &  1  &  9  &  7  & & 2 &  17 & 14  & 13  \\
\cline{1-4} \cline{6-9}
3 & 1 & 3 & 0.2  & & 3 & 6  & 0.1 & 0.08  \\
\cline{1-4} \cline{6-9}
4 & 2  & 11 & 2 & & 4 & 17 & 4 & 4 \\
\cline{1-4} \cline{6-9}

\end{tabular}
\end{table}

I also compare with the fraction of ground states with a given $J$ predicted by my 
simple random matrix picture, $f_\mathrm{RM}$. The only input are the dimensions $N_J$
 and the universal variances computed in
(\ref{SU2variance}) and used in (\ref{scaledwidth}). 

For such a simple picture, the random matrix model yields qualitatively 
excellent results, generally predicting the enhancement or suppression of different 
$J$s in the CI simulations relative to the native fractions $f_\mathrm{space}$. 
In particular, not only does the RM model successfully predict an enormous enhancement 
of $J=0$ in the ground state, it predicts a mild enhancement of $J=2$.

This analysis suggests the predominance of angular-momentum zero 
ground states is primarily a function of the width of the angular-momentum-projected 
many-body Hamiltonian; furthermore, the width is largely decoupled from the 
microphysics, instead depending only on the projection integrand  (\ref{SU2variance}) 
and on the dimensionality of subspaces with good quantum numbers. 
The simplicity and decoupling from the microphysics may be why the phenomenon is 
so robust and so universal.

So far I have only considered angular momentum. Yet 
in nature, nuclei with even numbers of protons and even numbers of neutrons 
always have $J=0$ ground states while those with odd numbers of 
protons and odd numbers of neutrons often do not, and if one runs ensembles of 
configuration-interaction simulations with two-body interactions that conserve 
angular momentum and and isospin, this scenario is broadly reproduced: one gets a 
predominance of $J=0$ ground states  for even-even cases but greatly reduced for 
odd-odd. 

\begin{table}
\caption{ Distribution of ground state angular momentum $J$ 
for systems with equal numbers of protons and neutrons, 
given as a percent 
for configuration-interaction simulations ($f_\mathrm{CI}$)
against fraction in the simple random-matrix model 
($f_\mathrm{RM}$). 
 The columns marked `conserved' means 
isospin consevation was enforced, while for those marked 
`broken' isospin was maximally broken. The single-particle 
model spaces are $1s$-$0d$ or $(sd)$ 
and $1p$-$0f$ or $(pf)$.
\label{isospin}}

\begin{tabular}{|r|r|r|r|r||r|r|r|}
\colrule
  &  \multicolumn{4}{l||}{ conserved } & \multicolumn{3}{|c|}{ broken } \\
\colrule
$J$ & $f_\mathrm{space}^{T=0}$ & $f_\mathrm{space}^{T=1}$ 
& $f_\mathrm{RM}$ &  $f_\mathrm{CI}$  & 
$f_\mathrm{space}$ & $f_\mathrm{RM}$  &  $f_\mathrm{CI}$  \\
\colrule
\multicolumn{8}{|c|}{ $(sd), Z=N=3$ } \\
\colrule
0  & 0.8  & 1.6  & 42 & 15 & 3.6 & 72  & 32 \\
\colrule
1 & 2.5  & 4.3 &   32 & 34  & 10 &  16  & 31 \\
\colrule
2  & 3.5 & 6.4  &  17  &  9  & 15 &  10 & 14  \\
\colrule
3  & 4.2 & 7.3 &    7 & 26 & 17&       2 & 15  \\
\colrule 
4  & 4.1  & 7.2 &   0.5  & 1.6 & 16 & .1  & 4 \\
\colrule
\multicolumn{8}{|c|}{ $(pf),Z=N = 3$ } \\
\colrule
0   & 0.7  & 1.2 & 41 & 11 & 2.6 & 66 & 28 \\
\colrule
1   & 2.1  &3.4  & 27 & 36 & 7.6 & 16  & 26 \\
\colrule
2   & 3.0  & 5.2  & 19  &  7  & 11 & 13 & 16  \\
\colrule
3  & 3.8 & 6.2 & 11 & 23 &  13 & 3 & 15  \\
\colrule 
4  & 3.9 & 6.6 & 2  & 2 & 14 & 0.5  & 3 \\
\colrule
\end{tabular}

\end{table}

I now generalize the above simple model, and consider widths that depend upon both 
total angular momentum $J$ and total isospin $T$. If $N_{JT}$ is the dimension of a CI 
space with fixed $J$, $T$, then let the width be 
\begin{equation}
\sigma^\mathrm{eff}_{JT} = \sqrt{N_{JT}} \sigma_J \sigma_T
\end{equation}
where $\sigma_T$ is also taken from (\ref{SU2variance}). 

Table \ref{isospin} shows the dimensions for several proton-neutron cases, 
both for fixed $J$, $T$ and for fixed $J$ alone, as well as the random matrix 
prediction for the fractions $f_{JT}$ and $f_{J}$.  For even-even cases (not shown 
to save space), there is 
little difference, but for odd-odd, breaking isospin dramatically enhances the fraction 
of $J=0$ ground states. This can be considered a prediction of the simple 
random matrix model. The results can be traced directly to the subspace dimensions 
(here $f_\mathrm{space}$). Specifically, as one goes from isospin breaking to 
isospin conserving, a smaller fraction of $J=0$ states go to $T=0$ than do 
$J=1$, and the relative decrease of dimensionality of $J=0,T=0$ makes 
the difference.

I include the results of an ensemble of CI simulations, which 
verify the qualitative predictions of the random matrix model. 
%(Intriguingly, although it is not shown, if use a charge-symmetric but isospin breaking interaction, that is $V_{pp} = V_{nn}$ but $V_{pn}$ is independent in all channels, the 
%enhancement of $J=0$ is even larger. This deserves further investigation.) 
While that quantitative agreement is significantly less than for the single-species 
case, the qualitative trends do agree. 
%(In future work I will look at specific coupling of proton and neutron angular momenta, which I expect to improve the results.)

%It would be useful to go beyond simply tying ground state quantum numbers to widths; 
%for example, can one learn something more about the ground state wavefunction? 
%Consider the fact that, for example, atomic nuclei typically have 
%two species of particles, protons and neutrons. I can generalize to have wavefunctions 
%that depend upon two angles, one for each species: $\psi(\theta_p \phi_p, \theta_n \phi_n)$.
%One expects--and in a moment I'll show it is true--that this wavefunction must 
%be expressible in terms of spherical harmonics, specifically 
%\begin{equation}
%\psi_{LM} = \sum_{L_p, L_n} c^L(L_p, L_n) \left [ Y_{L_p}(\theta_p \phi_p) 
%\otimes Y_{L_n}(\theta_n \phi_n) \right ]_{LM}.
%\end{equation}
%Can we learn anything about the distribution of the $c^L(L_p,L_n)$ for the 
%ground state wavefunction? The obvious suspicion is that it, too, would be dominated 
%by low quantum numbers. Once again, I can derive this using projection integrals on 
%random matrices, and also explicitly with many-body simulations. 

%(Interestingly, if I take the scaling in (\ref{scaledwidth}) with $N_J$ to be 
%slightly different--say $N_J^{0.6}$ or $N_J^{0.4}$ I get radically different results, 
%for example, an $J=0$ fraction as being $\sim 0.3$ or $1.0$, respectively.)

\begin{figure}  
\includegraphics [width = 7.5cm]{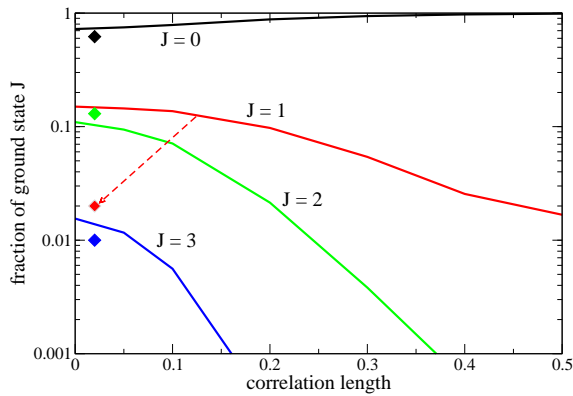}
\caption{\label{correlationfig} (color online) 
Fraction of ground states with angular momentum $J$ as a function 
of the correlation length (see Eq.~\ref{correlatedwidth}), for 
$N=Z=4$ in the $sd$ shell. Diamonds correspond to the TBRE result; only the $J=1$ is shifted significantly,
see dashed arrow.}
\end{figure}

In computing (\ref{SU2variance}), I assumed each 
$\mathbf{F}(\omega)$ to be uncorrelated, so that the widths add incoherently. 
Suppose instead that  that $\mathbf{F}(\omega)$ and $F(\omega^\prime)$ are correlated, 
with a correlation length $\gamma$; then the width (\ref{SU2variance}) becomes 
\begin{eqnarray}
4\pi^2 \bar{\sigma}^2 \int_0^\pi d\omega \int_0^\pi d\omega^\prime 
P_J(\cos \omega) \sin \omega P_J(\cos \omega^\prime) \sin \omega^\prime 
 \nonumber \\ \times \exp \left ( -| \omega-\omega^\prime  |/\gamma
\right ). \label{correlatedwidth}
\end{eqnarray}
In the limit $\gamma \rightarrow \infty$, which corresponds for a perfectly 
correlated, i.e. constant, $\mathbf{F}$ this vanishes for $J>0$.  
Fig.~(\ref{correlationfig}) shows how the fraction of ground states with $J$ 
varies with correlations length $\gamma$, for the case of $^{24}$Mg.

Even if $\mathbf{F}(\omega)$ are not correlated, the $\mathbf{h}_J$ still have 
correlations because 
$\int P_J(\cos \omega) P_{J^\prime}(\cos \omega) \sin^2 \omega d\omega \neq 0$; this 
correlation was found via a much more complicated prior analysis \cite{PW04}.   One immediate result follows from the integral: while $J=0$ is 
correlated with all even $J$, it is not correlated with odd $J$. Further consequences 
will be pursued in future work.

%  CONCLUSION

In summary, detailed quantum many-body simulations using ensembles of random interactions have shown 
surprising trends, most notably the predominance of angular momentum $J=0$ in the ground state. %Because 
%this predominance is robust, 
%we should look beyond details 
%of the microphysics for the answer. 
To address this question, I have argued how one can project 
random matrices with good quantum numbers. 
Using only simple, universal integrals and the dimensionality of subspaces with 
good quantum numbers, I can qualitatively reproduce the features of ensembles of 
many-body systems. In particular I find an dominance of $J=0$ for the 
ground state, although I reproduce other trends as well.  When I further consider systems with 
two species (protons and neutrons) the random matrix model predicts, and CI simulations 
confirm, that the subspace dimensions of 
isospin-conserving systems suppresses $J=0$ ground states in  
odd-$Z$, odd-$N$ systems. 

The results of the model are qualitative, not quantitative. On the other hand, 
the effectiveness of the qualitative results for a broad variety of cases suggests 
that some properties of many-body systems are founded not in detailed microphysics 
but on simple properties, specifically projection integrals 
and the relative dimensions of subspaces with good quantum numbers.

%It will be 
%interesting to see what further insights  can be gleaned using this approach. 
%Further investigation would be particularly warranted into symmetry breaking, two-species 
%systems, and correlations between states with different quantum numbers. 

The U.S.~Department of Energy supported this investigation through
grant DE-FG02-96ER40985. 
% I thank Erich Ormand for installing isospin-breaking capability into the BIGSTICK code. 

\end{document}